\documentclass[conference]{IEEEtran}
\IEEEoverridecommandlockouts

\usepackage{xspace}
\usepackage{amsmath,amssymb,amsfonts}
\usepackage{algorithmic}
\usepackage{graphicx}
\usepackage{textcomp}
\usepackage{xcolor}
\usepackage{xspace}
\usepackage{numprint}
\usepackage{nicefrac}
\usepackage{multirow}
\usepackage{caption}
\usepackage{hyperref}
\usepackage{paralist} 
\usepackage{enumitem}
\usepackage{subcaption}
\def\BibTeX{{\rm B\kern-.05em{\sc i\kern-.025em b}\kern-.08em
    T\kern-.1667em\lower.7ex\hbox{E}\kern-.125emX}}
\DeclareMathAlphabet{\mathcal}{OMS}{cmsy}{m}{n}
\newcommand{\ournameNoSpace}{FLAIRS}
\newcommand{\ourname}{\ournameNoSpace\xspace}
\newcommand{\ournameGen}{\ournameNoSpace's\xspace}
\newcommand{\paperTitle}{\ourname: FPGA-Accelerated Inference-Resistant \& Secure Federated Learning }

\newcommand{\adversaryNoSpace}{\ensuremath{\mathcal{A}}}
\newcommand{\adversary}{\adversaryNoSpace\xspace}

\newcommand{\adversaryServerNoSpace}{\ensuremath{\mathcal{A^S}}}
\newcommand{\adversaryServer}{\adversaryServerNoSpace\xspace}

\newcommand{\etal}{\emph{et~al.}}
\newcommand{\sect}{Sect.~}

\newcommand{\numberOfMaliciousClients}{\ensuremath{n_\adversary}\xspace}

\newcommand{\hdbscanPE}{HDBSCAN PE\xspace}
\newcommand{\cifar}{CIFAR-10\xspace}

\newcommand{\flame}{FLAME\xspace}
\newcommand{\flameGen}{FLAME's\xspace}

\newcommand{\scaleTable}[1]{\scalebox{.825}{#1}}

\definecolor{darkgreen}{rgb}{0.0, 0.5, 0.0}

\definecolor{purple}{rgb}{0.5, 0.0, 0.5}

\usepackage{tikz}
\newcommand*\circled[1]{\tikz[baseline=(char.base)]{
            \node[shape=circle,draw,inner sep=1pt] (char) {#1};}}
            
\begin{document}

\title{\paperTitle}

\author{
    \IEEEauthorblockN{Huimin Li\IEEEauthorrefmark{1}, Phillip Rieger\IEEEauthorrefmark{2}, Shaza Zeitouni\IEEEauthorrefmark{2}, Stjepan Picek\IEEEauthorrefmark{3}\IEEEauthorrefmark{1} and Ahmad-Reza Sadeghi\IEEEauthorrefmark{2}}
    \IEEEauthorblockA{\IEEEauthorrefmark{1}Delft University of Technology, The Netherlands, H.Li-7@tudelft.nl}
    \IEEEauthorblockA{\IEEEauthorrefmark{2}Technische Universit\"at Darmstadt, Germany, \{phillip.rieger, shaza.zeitouni, ahmad.sadeghi\}@trust.tu-darmstadt.de}
    \IEEEauthorblockA{\IEEEauthorrefmark{3}Radboud University, The Netherlands, stjepan@computer.org} 
}

\maketitle

\begin{abstract}
Federated Learning (FL) has become very popular since it enables clients to train a joint model collaboratively without sharing their private data. However, FL has been shown to be susceptible to backdoor and inference attacks. While in the former, the adversary injects manipulated updates into the aggregation process; the latter leverages clients' local models to deduce their private data. Contemporary solutions to address the security concerns of FL are either impractical for real-world deployment due to high-performance overheads or are tailored towards addressing specific threats, for instance, privacy-preserving aggregation or backdoor defenses.
 
Given these limitations, our research delves into the advantages of harnessing the FPGA-based computing paradigm to overcome performance bottlenecks of software-only solutions while mitigating backdoor and inference attacks. We utilize FPGA-based enclaves to address inference attacks during the aggregation process of FL. 
We adopt an advanced backdoor-aware aggregation algorithm on the FPGA to counter backdoor attacks. We implemented and evaluated  our method on Xilinx VMK-180, yielding a significant speed-up of around 300 times on the IoT-Traffic dataset and more than 506 times on the CIFAR-10 dataset.

\end{abstract}


\section{Introduction}
\label{sec:intro}
FPGAs are powerful and versatile devices, providing flexible platforms for custom hardware solutions. With unique characteristics like parallel processing, support for various data types, low latency, and lower power consumption compared to general-purpose computing platforms, they excel in accelerating computations and tackling complex challenges. FPGAs have become indispensable across various domains, from high-performance computing and data centers to the Internet of Things (IoT) and embedded systems. Their widespread adoption is evident in their deployment within commercial cloud platforms such as Amazon EC2~\cite{amazon-f1}, Microsoft Azure~\cite{catapult}, and Alibaba Cloud~\cite{alibaba}, underscoring their significance and impact in today's technological landscape.

In addition to their inherent benefits, FPGAs enable establishing a trusted execution environment (TEE), ensuring the security of critical workloads, including the FPGA configuration, which may comprise an Intellectual Property (IP) design, and the processed data, without compromising performance.  Recent advancements in FPGA research have demonstrated the feasibility of establishing TEEs on commodity FPGAs deployed in cloud environments~\cite{fpgatee21,ZGK22}. Consequently, FPGAs can provide not only acceleration but also secure processing of clients' workloads in hostile cloud environments. 
This paradigm shift toward trusted execution on cloud FPGAs offers numerous advantages. It grants organizations greater control over their applications and data security, even when physical access to the FPGA is limited or non-existent. We refer to TEEs on FPGAs as FPGA-based TEEs to distinguish them from TEEs on CPUs. 

\textbf{Federated Learning.} One of the compelling applications for FPGA-based TEEs is Federated Learning (FL), a collaborative learning approach.
Unlike traditional centralized learning, FL allows clients to train their own DNN models locally using their private datasets and share only the training results with a central server that aggregates clients' models or local models into a global model and returns it to the clients. Hence, sensitive data remains confined to clients' computing premises. Therefore, FL is foreseen to improve the clients' privacy \cite{mcmahan2017}. 

However, FL is still prone to privacy attacks during the aggregation process. Such attacks aim to infer information about the training data of a model, e.g., if a specific sample was used~\cite{shokri} or try to reconstruct samples from the training data \cite{salem2020updates}. Although individual clients' contributions are anonymous, preventing associating the inferred information with a specific client, a malicious aggregation server can still exploit access to the local models to analyze them and violate clients' privacy.

Another type of attack on FL targets the model's integrity. Poisoning attacks aim to manipulate the global model to misbehave, i.e., \textit{targeted or backdoor attacks}~\cite{bagdasaryan2020how,shen16Auror,nguyen2020diss}, or they aim at rendering the model useless, i.e., \textit{untargeted attacks}~\cite{shejwalkar2021manipulating,fang2020local}. Targeted attacks are crucial because the adversary injects a stealthy function to influence the outcome without violating the model's utility. 

\textbf{Existing Defenses.} Proposed defenses for FL typically address only one type of attack, focusing either on protecting clients' privacy~\cite{bonawitz,fereidooni2021safelearn,wang2022pipefl} against a potentially malicious server, or mitigating specific backdoor attacks launched by malicious clients~\cite{shen16Auror,blanchard17krum,jebreel2023fl,rieger2022deepsight}. 
Mitigating both types of attacks poses a dilemma. On the one hand, detecting and filtering poisoned models requires the aggregator to inspect local models. On the other hand, privacy defenses prevent the aggregator from inspecting the local models. This presents a challenge in striking the right balance between security and privacy in FL presents a challenge.

To solve this dilemma, several privacy-preserving approaches such as Homomorphic Encryption (HE) or Secure Multi-Party Computation (SMPC)~\cite{khazbak2020mlguard,nguyen22Flame,tian2022flvoogd} have been proposed to process models without divulging any information. However, such solutions result in significant performance overhead and scalability issues, particularly for complex backdoor defenses involving vector metric computations or clustering. Implementing these defenses using SMPC, such as in the case of~\cite{nguyen22Flame,khazbak2020mlguard}, becomes highly impractical due to the associated overhead.
An alternative approach is to utilize TEEs on CPUs to ensure local models' privacy while the aggregator inspects them. For instance, TEE-based implementations of Krum \cite{blanchard17krum} have been explored \cite{mondal2021poster,hashemi2021byzantine}. However, TEEs' limited computation capacities introduce significant overhead for computation-intensive algorithms like Krum, which involves calculating Euclidean distances between local models. 

Therefore, utilizing FPGA-based TEEs seems to be an intuitive approach for achieving secure and privacy-preserving FL.
Among recent software-based proposals, \flame aims to address backdoor and inference attacks to be independent of the attack strategy. To counter backdoor attacks, \flame combines outlier-detection-based filtering with model clipping and noising. However, \flame suffers from significant performance overhead due to the deployment of SMPC for protecting clients' privacy. 
SMPC protocols enable the secure evaluation of a public function, e.g., the aggregation process, on private data, e.g., local models, from $N$ mutually distrusting parties. SMPC finds utility in outsourcing scenarios \cite{demmler2015aby}, where multiple parties/clients can secret-share their private inputs among two or more non-colluding, well-connected, and powerful servers responsible for executing the SMPC protocol, yet resulting in a significant computation overhead and hence does not scale. In the case of \flame, for aggregating 50 models trained on the \cifar dataset, SMPC increases the execution time of \flame from $2.6$s to $766.1$s. Additionally, SMPC requires non-colluding aggregation servers. Consequently, the privacy guarantees of \flame only apply to semi-honest aggregation servers that adhere to the SMPC protocol \cite{nguyen22Flame,demmler2015aby}.  

\noindent\textbf{Goals and Contributions.} In this work, we propose to leverage FPGA-based TEEs to enable privacy-preserving backdoor-aware aggregation for FL. Our optimized FPGA-accelerated approach enables the aggregation server to perform a privacy-preserving backdoor analysis of the local models with only a negligible computation overhead. The described techniques allow the acceleration of arbitrary backdoor defenses. We exemplary prototype our approach using the recently proposed defense \flame~\cite{nguyen22Flame}.

Our contributions can be summarized as follows:
\begin{itemize}
\item We leverage FPGA-based TEEs, demonstrating a practical and efficient backdoor-aware FL aggregation while protecting clients' privacy in a stronger adversary model than SMPC. 
\item Our approach generally allows implementing arbitrary backdoor resilient aggregation schemes on secure FPGAs.
\item We demonstrate the security and performance gain of \ourname by realizing exemplary the entire \flame~\cite{nguyen22Flame} algorithm on FPGA, resulting in speed-ups of over 288 times on the IoT-Traffic dataset and more than 506 times on the \cifar dataset. The above results are obtained from a single FPGA. However, the runtime can be reduced by almost $m$ if $m$ FPGAs run in parallel.
\item We propose the cascade structure, enabling the calculation of cosine distance with time complexity of $O(n)$ instead of $O(n^2)$. This structure also significantly reduces the access time to the main memory.

\end{itemize}

\section{Background}
\label{sec:background}

\textbf{Remote Attestation} verifies the authenticity and integrity of code or memory on a remote device. The \emph{verifier} receives a signed cryptographic hash of the content from a trusted component on the \emph{prover} device. The received digest is compared against the expected reference value to determine the prover's status.

\textbf{Trusted Execution Environments (TEEs)} 
provide secure enclave applications isolated from all other untrusted software in the system. The TEEs aim to protect the confidentiality and integrity of the enclave's code and data. 
Remote attestation can be used by clients before sharing confidential data with enclaves. Code and data are processed unencrypted in the CPU's caches and registers but are encrypted and integrity-protected with an enclave-specific secret key before storing in untrusted storage. Commercial TEEs like Intel SGX~\cite{sgx}, AMD SEV~\cite{sev}, and ARM TrustZone~\cite{trustzone} are widely deployed in computing systems, including FPGA-SoCs such as ARM TrustZone on Intel Stratix 10 SoC and AMD Xilinx ZCU102.
\noindent\textbf{TEEs on FPGAs.} 
FPGA trusted execution protects IP configuration and processed data. FPGA manufacturers like Intel and AMD Xilinx offer hardened cryptographic cores for IP confidentiality and integrity. For cloud deployments without physical access, a trusted third party or vendor is necessary~\cite{EguVen12,HKKT17,EAA19,fpgatee21}. Thereafter, two methods of establishing FPGA TEEs exist (1) loading IP design on SoCs with built-in TEEs to guarantee FPGA trustworthiness \cite{KNLB21,ZGK22}; and (2) using a trust anchor (secure shell) for FPGAs without TEEs~\cite{fpgatee21} which provides remote key generation, configuration, isolation, and cryptographic operations. Here, remote attestation ensures the authenticity and integrity of the FPGA configuration~\cite{VRCM19,fpgatee21}.

\section{Problem Setting}
\label{sec:problem}

\begin{figure}[t]
    \begin{center}
    \includegraphics[width=0.7\linewidth,trim={6.65cm 9.25cm 7.8cm 1.cm},clip]{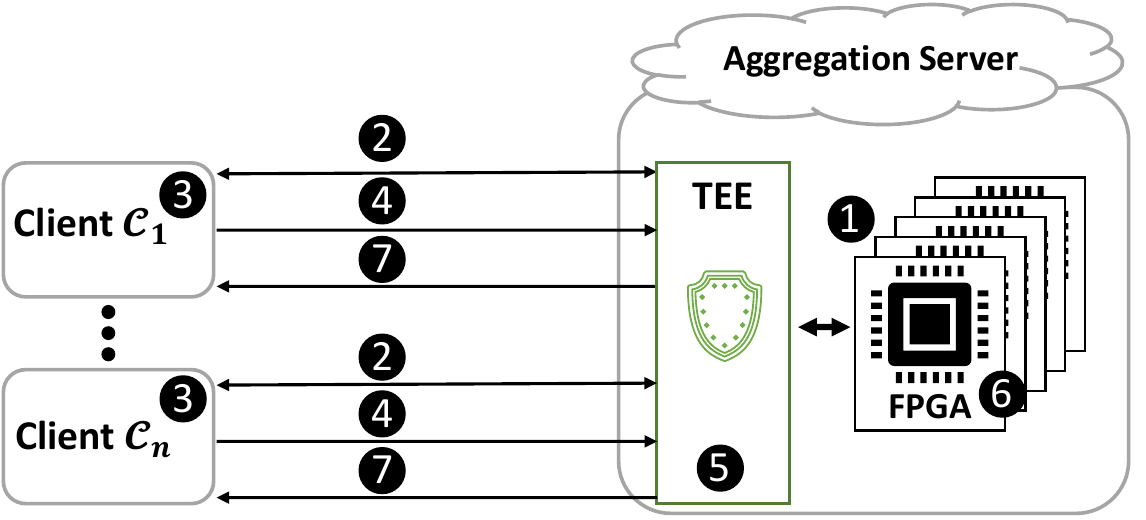}
    \captionsetup{justification=centering}
    \caption{Workflow of \ourname.}
    \label{fig:sysmodel}\vspace{-0.3cm}
    \end{center}
\end{figure}

\subsection{System \& Adversary Model}
\label{sec:problem-threat}
We consider a system consisting of an aggregation server and $n$ clients jointly training a DNN model. Each client $\mathcal{C}_i$ trains locally a model with its dataset and sends the aggregator its local model. 
The aggregation runs on the cloud and can therefore benefit from one or multiple FPGAs to accelerate the aggregation.
The system setting is visualized in Fig.~\ref{fig:sysmodel}, together with the individual steps of \ourname, which will be described in \sect\ref{sec:problem-design}.

We consider two types of adversaries: (1) \adversary that aims to inject a backdoor, and (2) \adversaryServer that aims to infer information about clients' training data.

To inject a backdoor, \adversary changes the predictions of all samples within a trigger set $\mathcal{I}\subset\mathcal{D}$ towards a specific label. \adversary must ensure that the attack is not detected, which includes preventing a significant drop in the aggregated model's utility on regular samples. We assume \adversary fully controls $\numberOfMaliciousClients<\nicefrac{n}{2}$ clients and can manipulate their training process and data. However, \adversary does not know the data or models of other clients.

On the malicious server side, aligned with existing work, \adversaryServer aims to extract information from the individual local models~\cite{andreina2020baffle,nguyen22Flame,khazbak2020mlguard} since the aggregation anonymizes the individual contributions of clients and prevents an adversary from associating information gained from the aggregated model with particular clients. We assume that \adversaryServer controls the aggregation server and a few clients, has full software-level access, and can arbitrarily deviate from the aggregation process. 
We exclude denial-of-service attacks intended to shut down computational resources, as they can be detected.
Moreover, we assume that physical attacks on the infrastructure, including the FPGAs, are out of scope. However, remote physical attacks performed using malicious FPGA configurations can be mitigated using FPGA scanners~\cite{KGT19,LMGP20}, as demonstrated in~\cite{fpgatee21}. As we show next, all clients can vet the FPGA configurations that represent the accelerators and verify their integrity and authenticity as a part of attesting the TEE. 

\subsection{\ourname Overview}
\label{sec:problem-design}
To achieve secure and practical backdoor-aware FL, we adapt \flame~\cite{nguyen22Flame} to run on a FPGA-based TEE \cite{EguVen12,HKKT17,EAA19,fpgatee21,KNLB21,ZGK22}.
Note that the entire aggregation algorithm (demonstrated in \sect\ref{subsec:ana_flame}) is unlikely to fit on a single FPGA, considering a large number of clients and model parameters. Therefore, the aggregation algorithm can be split into several accelerators and benefit from using several FPGAs or swapping in and out the different accelerators on a single FPGA. Hence, when multiple FPGAs/accelerators are used, a scheduler algorithm must coordinate the work of the accelerators and receive clients' models. The scheduler can be implemented as a software application in a TEE or a hardware IP continuously running on the FPGA. In both cases, the scheduler can be attested by clients to ensure its authenticity and integrity.
In the following, we describe \ournameGen workflow (Fig.~\ref{fig:sysmodel}). 

\noindent\textbf{Step~\circled{\small{1}}.} This step establishes a TEE on the cloud FPGA, where clients' models can be processed securely \cite{fpgatee21,ZGK22}.

\noindent\textbf{Step~\circled{\small{2}}.} The clients attest the TEE, i.e., verify the integrity and authenticity of the FPGA configurations that process clients' models. 
Thus, the clients have the assurance that the code processing their models is benign, i.e., not corrupted by \adversaryServer, and can exchange secret session keys with the TEE to encrypt their models.

\noindent\textbf{Step~\circled{\small{3}}.} The clients encrypt their models using individual secret session keys exchanged with the TEE.

\noindent\textbf{Step~\circled{\small{4}}.} The clients send their encrypted models to the aggregator.

\noindent\textbf{Step~\circled{\small{5}}.} 
The models are then stored in memory, ready for aggregation.

\noindent\textbf{Step~\circled{\small{6}}.} The TEE and FPGAs use \flame to aggregate the models and mitigate backdoor attacks.

\noindent\textbf{Step~\circled{\small{7}}.} The aggregated model is sent back to all clients.  

\section{Design \& Implementation}
\label{sec:implement}

\subsection{Analysis of \flame Algorithm}
\label{subsec:ana_flame}

\begin{figure}[b]
    \centering  \includegraphics[width=0.9\linewidth]{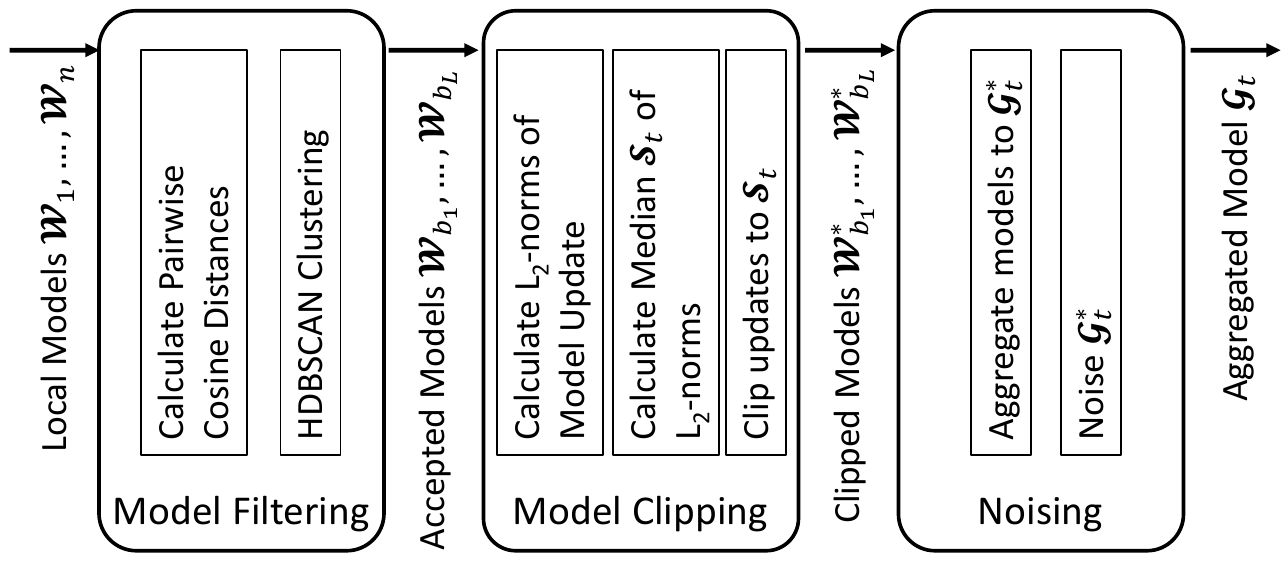}
    \caption{High-Level Overview of \flame \cite{nguyen22Flame}}
    \label{fig:flame}
\end{figure}

We adopt \flame \cite{nguyen22Flame} for backdoor-aware aggregation, which consists of three defense layers, namely Model Filtering, Model Clipping, and Noising, shown in Fig.~\ref{fig:flame}. Adding noise can remove the backdoor but will also drop the models' utility in terms of accuracy on the main task. While filtering and clipping decrease the amount of noise required to mitigate poisoned models. We thoroughly analyze \flame and define efficient hardware components or processing elements (PEs).

To optimize \flame performance, we break down the compute-intensive cosine distance (Model Filtering) into two parallel components: preprocessor and cosine similarity.
\textbf{Preprocessor} component $Prep.\ PE$ calculates the differential vector and Euclidean distances $L_2\_norms$ between local models and the global model.
\textbf{Cosine-similarity} component $Cosine PE$ runs parallel to $Prep.\ PE$ and computes the dot products of clients' differential vectors, which, along with $L_2\_norms$, are used to calculate cosine distances for all local models. 
\textbf{HDBSCAN} (Model Filtering) labels local models based on cosine distances as benign models $1$ or malicious models $0$.
\textbf{Scale} (Model Clipping) calculates the median value ($S_t$) of $L_2\_norms$ and generates scaled models for all local models. HDBSCAN and Scale PEs run in parallel with no data dependency.
The \textbf{Aggregation} $Agg.\ PE$ comprises clipping (Model Clipping), aggregation, and noise addition (Nosing) steps. It generates the final aggregated model by using models' labels, median values, and model scales from previous components.

\begin{figure}[tb]
    \centering
\includegraphics[width=0.9\linewidth]{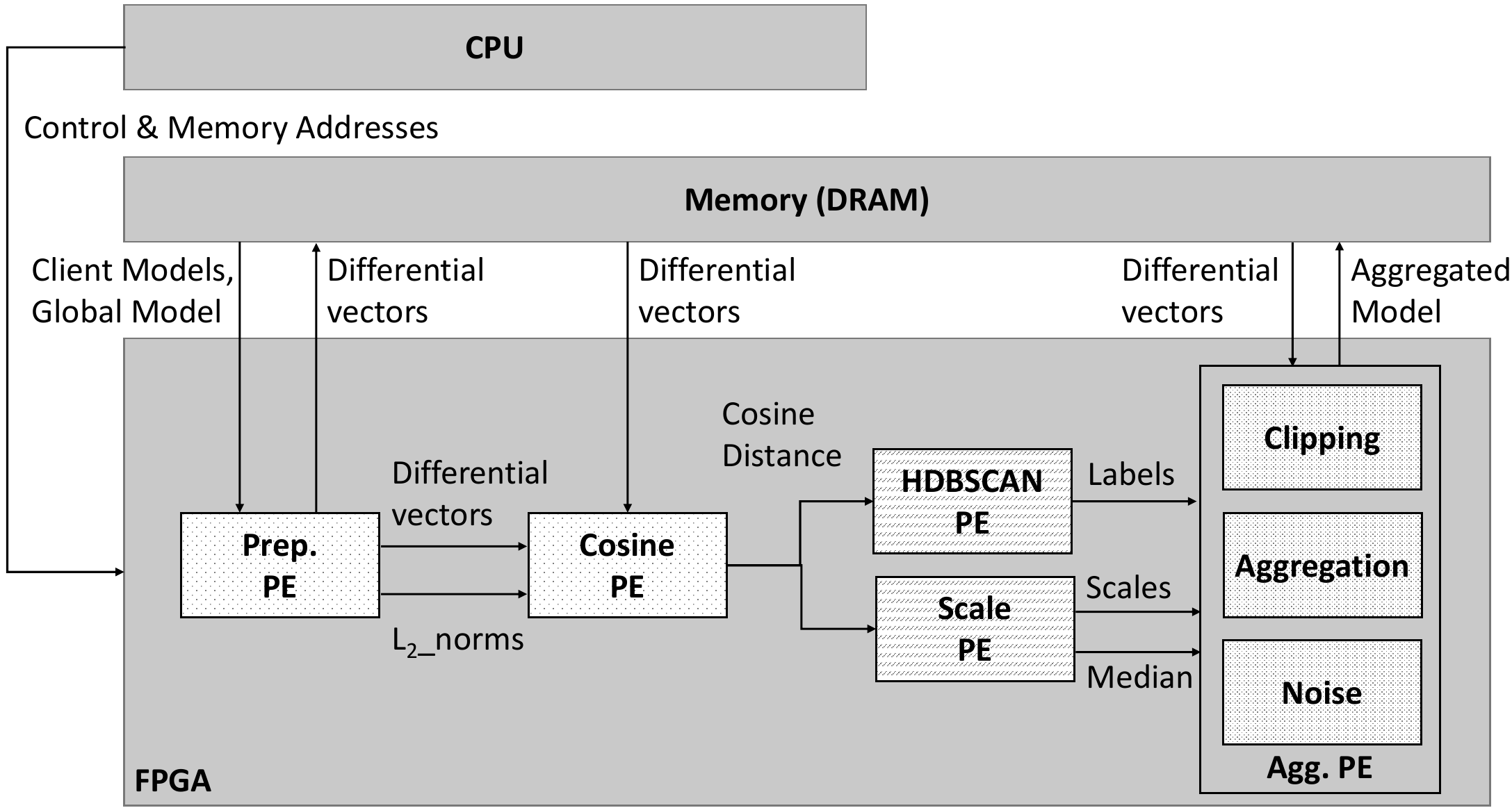}
    \caption{System Architecture of \ourname}
    \label{fig:sys fpga}
\end{figure}

\subsection{Implementation}
\label{subsec:impl}
We implemented \ourname (Fig.~\ref{fig:sys fpga}) on Xilinx Vitis 2022.2 using the VMK180 Evaluation Kit, which has a built-in ARM TrustZone in its hardened processing unit. 
The system comprises a host program running on a TEE-enabled CPU and an FPGA platform consisting of \texttt{shell} and \texttt{kernel} components. The host program manages the FPGA's components and the operation process~\cite{xilinxDocUG1393}. The \texttt{shell} provides the essential functions for execution, security, and communication interfaces, while the \texttt{kernel} is the dynamic region for custom logic implementation. We utilized HLS to translate the C++ kernel module into device logic fabric and RAM/DSP block~\cite{xilinxDocUG1399}.
The operating frequency of the \texttt{kernel} is set to 300 MHz, and a burst mode was adopted to utilize AXI4 interface throughput fully. The width of the AXI4 Memory Mapped interface was configured to 512 bits, and the burst length is set to transfer 4KB each time \cite{xilinxDocUG1399}.

\subsubsection{Preprocessor PE}
\label{subsec:impl_pre}

The Preprocessor PE stores the global model locally on the FPGA, followed by the sequential transmission of local models from DDR-RAM to FPGA. Each local model undergoes subtraction with the global model to obtain its differential vector, which is then routed to the $Cosine PE$ and back to DDR-RAM. After that, each differential value is squared and accumulated, and the accumulated value is processed using the square root function to yield the $L_2\_norms$ value for each client. This process operates parallel within a pipeline structure defined by the total number of clients and parameters per local model.

\subsubsection{Cosine-similarity PE}
\label{subsec:impl_cosine}
The pairwise cosine distances can be represented as a matrix.  
The values on the diagonal are equal to $0$, while the remaining positions are determined by Eq.~\eqref{equ:dist}, where $d$ denotes the differential vector and $p$ represents the total number of parameters. The denominator of Eq.~\eqref{equ:dist} is obtained from $L_2\_norms$, and the numerator is derived from the dot product of two differential vectors of clients $i$ and $j$.
The cosine distance matrix is symmetric, so we only compute the upper triangular or lower part.
{\small{
\begin{equation}
\label{equ:dist}
dist_{i j}=1-\frac{d_i d_j}{\left\|d_i\right\|\left\|d_j\right\|}=1-\frac{\sum_{k=1}^{p} d_i^k d_j^k}{\sqrt{\sum_{k=1}^{p}\left(d_i^k\right)^2} \sqrt{\sum_{k=1}^{p}\left(d_j^k\right)^2}}
\end{equation}}}

\begin{figure}
     \centering
     \begin{subfigure}[b]{0.4\textwidth}
         \centering
         \includegraphics[width=\textwidth]{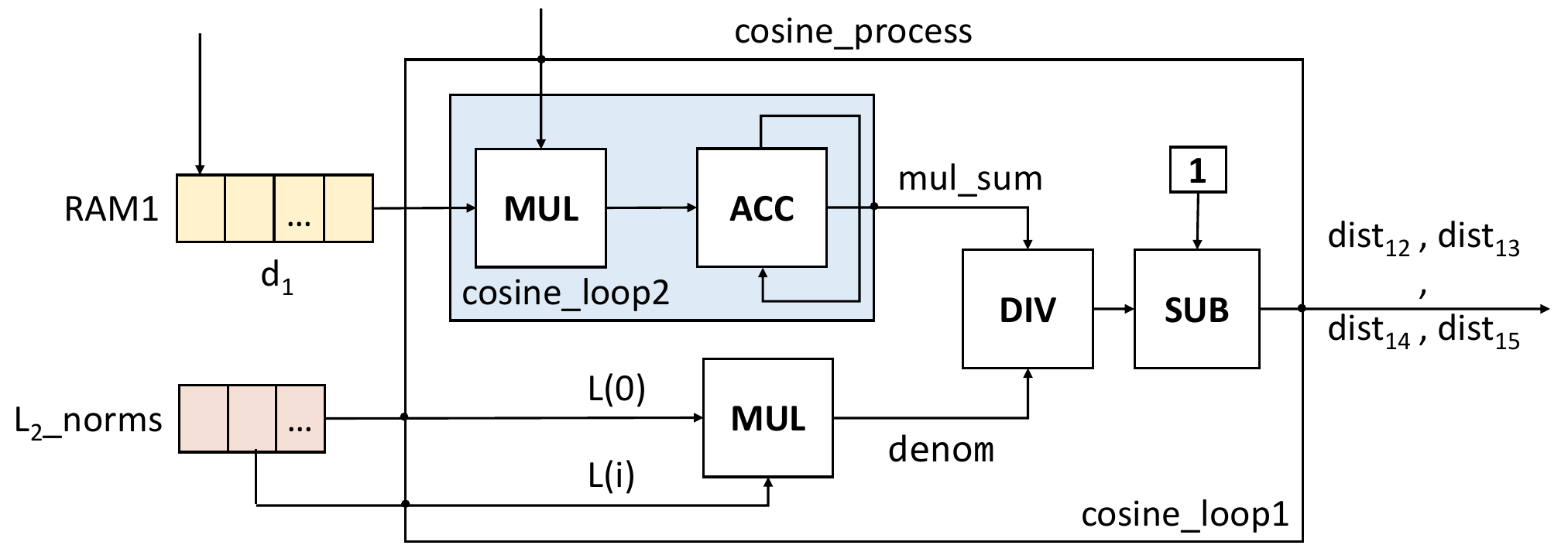}
         \caption{Cosine Process.}
         \label{fig:cosine}
     \end{subfigure}
     \hfill
     \begin{subfigure}[b]{0.4\textwidth}
         \centering
         \includegraphics[width=\textwidth]{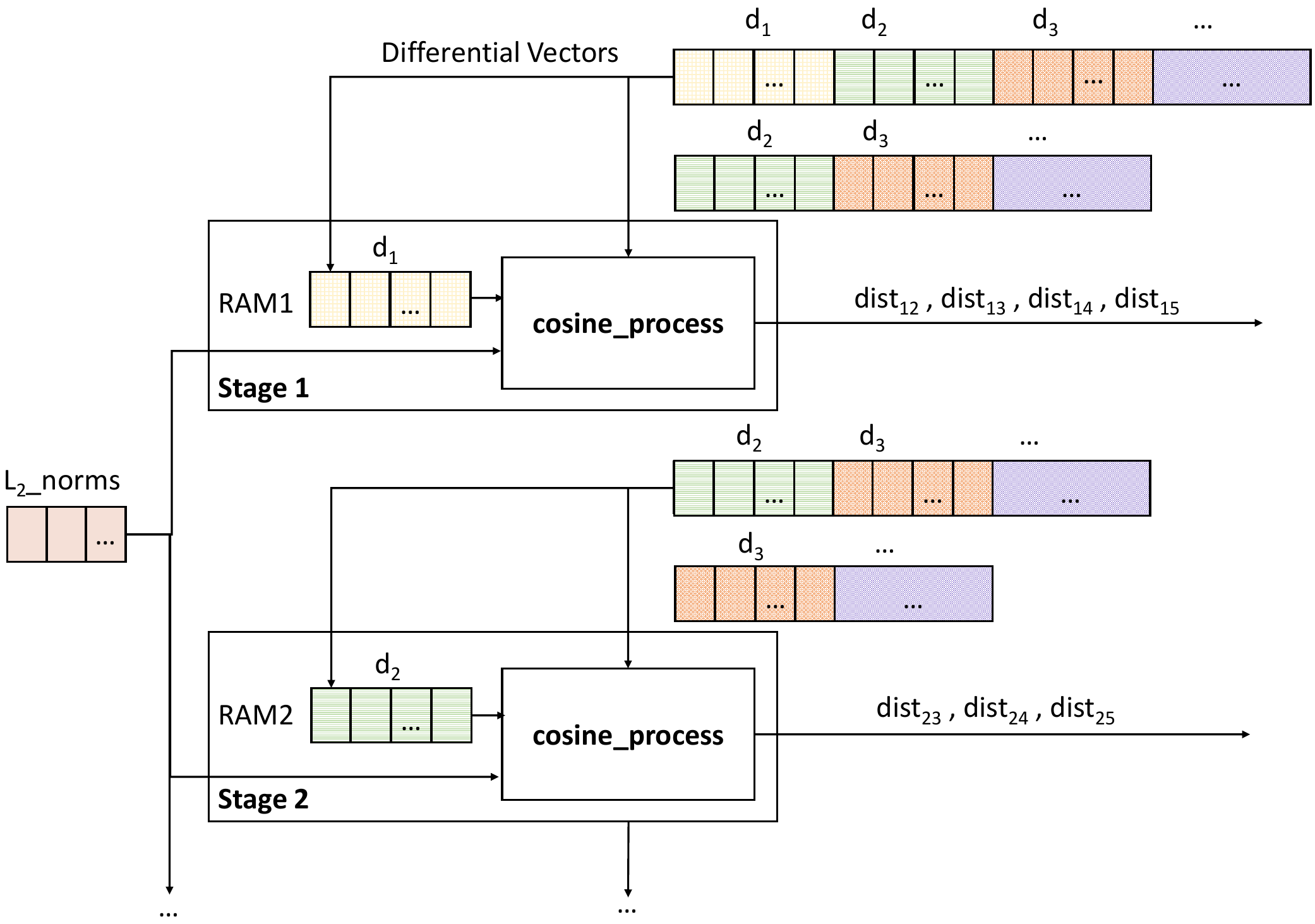}
         \caption{The Cascade Structure of Cosine-similarity.}
         \label{fig:Cascad structure}
     \end{subfigure}
    \caption{Cosine-similarity PE.}
\end{figure}

We use a cascade structure (Fig.~\ref{fig:Cascad structure}) to obtain dot products with multiple stages, where each stage locally stores the first-arriving differential vector in RAM. The cosine distance is calculated for the entire row using this initial vector and subsequent differential vectors in the $cosine\_process$ phase (Fig. \ref{fig:cosine}). 
Each stage, except the last one, sends out differential vectors (excluding the first vector) to its subsequent stage. 
The total number of stages depends on the device resources, number of clients, and number of parameters. If the number of stages is insufficient to calculate cosine distance for all local models, we utilize $hls::burst\_maxi{<}{>}$ to manually read the remaining differential vectors in bursts from DDR-RAM after each operation of the cascade structure. The new differential vectors are then processed through the same cascade structure for further cosine distance computations.

\begin{table*}[htb]
\caption{Runtime in seconds of FLAME using \ourname (\textbf{F}) compared to FLAME using SMPC (\textbf{S}) for \textbf{n} clients.}
\centering
\scaleTable{
\begin{tabular}{|c|c|c|cc|c|c|c|c|cc|c|}
\hline
\multirow{2}{*}{\textbf{Dataset}} &
  \multirow{2}{*}{\textbf{n}} &
  \textbf{Cosine Distance} &
  \multicolumn{2}{c|}{\textbf{HDBSCAN}} &
  \textbf{Scale} &
  \textbf{\begin{tabular}[c]{@{}c@{}}Aggregation\\      ( + Clipping +Noise)\end{tabular}} &
  \textbf{\begin{tabular}[c]{@{}c@{}}Kernel \\    Runtime\end{tabular}} &
  \textbf{\begin{tabular}[c]{@{}c@{}}Data   Transfer\\      (Host$\leftrightarrow$ DDR)\end{tabular}} &
  \multicolumn{2}{c|}{\textbf{Runtime}} &
  \multirow{2}{*}{\textbf{Speed-up}} \\ \cline{3-11}
 &
   &
  \textbf{F} &
  \multicolumn{1}{c|}{\textbf{F}} &
  \textbf{S} &
  \textbf{F} &
  \textbf{F} &
  \textbf{F} &
  \textbf{F} &
  \multicolumn{1}{c|}{\textbf{F}} &
  \textbf{S} &
   \\ \hline
\multirow{3}{*}{\textbf{IoT-Traffic}} &
  \textbf{10} &
  $5.3553   \times 10^{-2}$ &
  \multicolumn{1}{c|}{$1.5628\times   10^{-5}$} &
  3.64 &
  $1.420\times   10^{-6}$ &
  $1.6465\times   10^{-2}$ &
  $7.0034   \times 10^{-2}$ &
  $1.0677\times   10^{-2}$ &
  \multicolumn{1}{c|}{$8.0711\times   10^{-2}$} &
  108.16 &
  $1\ 340.1$ \\ \cline{2-12} 
 &
  \textbf{50} &
  0.453 &
  \multicolumn{1}{c|}{$1.509\times   10^{-3}$} &
  41.84 &
  $1.4019\times   10^{-5}$ &
  0.238 &
  0.693 &
  $1.1809\times 10^{-2}$ &
  \multicolumn{1}{c|}{0.705} &
  269.35 &
  382.1 \\ \cline{2-12} 
 &
  \textbf{100} &
  2.072 &
  \multicolumn{1}{c|}{$1.1851\times   10^{-2}$} &
  253.87 &
  $5.2265\times   10^{-5}$ &
  0.939 &
  3.023 &
  $1.2245\times   10^{-2}$ &
  \multicolumn{1}{c|}{3.035} &
  876.96 &
  288.9 \\ \hline
\multirow{2}{*}{\textbf{CIFAR-10}} &
  \textbf{10} &
  0.21 &
  \multicolumn{1}{c|}{$1.5628\times   10^{-5}$} &
  3.64 &
  $1.420\times   10^{-6}$ &
  $4.0984\times   10^{-2}$ &
  0.251 &
  $1.2104\times   10^{-2}$ &
  \multicolumn{1}{c|}{0.263} &
  134.93 &
  513 \\ \cline{2-12} 
 &
  \textbf{50} &
  1.235 &
  \multicolumn{1}{c|}{$1.509\times   10^{-3}$} &
  41.84 &
  $1.4019\times   10^{-5}$ &
  0.263 &
  1.5 &
  $1.2258\times   10^{-2}$ &
  \multicolumn{1}{c|}{1.512} &
  766.12 &
  506.7 \\ \hline
\end{tabular}
}
\label{tab:run_flame}
\end{table*}

\subsubsection{\hdbscanPE}
\label{subsec:impl_hdbscan}

\hdbscanPE determines one cluster representing the majority of models, thus containing at least $\nicefrac{n}{2}+1$, the minimum number of benign local models. 
The remaining models not part of this cluster are then considered noise.
For clustering, \flame uses HDBSCAN based on the implementation of McInnes \etal~\cite{mcinnes2017hdbscan}. 
Based on the \flameGen parametrization of HDBSCAN, a simplified version is implemented in the \hdbscanPE, neglecting all unused aspects. For example, there is no need to determine whether two closely packed dense groups of models constitute separate clusters or a single cluster.

\subsubsection{Scale PE}
\label{subsec:impl_scale}
Here, we first sort $L_2\_norm$ values from smallest to largest and then select the median value ($S_t$) from the sorted list. Subsequently, we calculate $\gamma[i] =\frac{S_t}{L_2\_norms[i]}$, where $i$ represents the client order. Finally, we obtain the model scale for each client as $min(1, \gamma[i])$.

\subsubsection{Aggregation PE}
\label{subsec:impl_agg}

In the aggregation PE, locally stored differential vectors from cascade structures are sequentially multiplied by corresponding scale values and added to the global model for models labeled as $1$ to get the resultant value $add\_sum$. If RAM does not have enough space to store all differential vectors, we use $hls::burst\_maxi{<}{>}$ to retrieve the remaining vectors from DDR-RAM. $add\_sum$ is accumulated for all local models, which is then divided by $accepted\_num$ (number of benign local models with label $1$) to generate a quotient. This quotient is added to the noise created using the $MT19937IcnRng$ function from Xilinx's Vitis Library~\cite{VitisLibrary}, generating random numbers following a normal distribution $N(0,1)$. The output of $MT19937IcnRng$ is multiplied with the required range $\lambda$ to conform to \flame's noise range.

\subsubsection{The Scheduler}
\label{subsec:impl_wrokflow}
 
The scheduler is the host program that runs on the TEE-enabled CPU and orchestrates the work of the FPGA accelerators, i.e., the kernels. 
Once the aggregation process is initialized, the scheduler is set as an enclave application. Clients attest the authenticity and integrity of the scheduler to ensure its code has not been modified. 
The scheduler then receives encrypted local models, decrypts \& re-encrypts them with a unified secret key, and stores them in the DDR-RAM.
The scheduler detects the Xilinx device, attests the FPGA binary file, and programs it into the device. Then, it creates the buffers the kernel needs in the DDR-RAM and sets up the kernel's input parameters (mapping ports).
Later, the scheduler writes data into buffers, triggers kernel execution, and waits for notification upon completion. Finally, it reads and sends the aggregated model to clients.

\subsection{Evaluation}
\label{subsec:eval}

To ease the comparison with \flame, we replicated the setup in \cite{nguyen22Flame} and evaluated \ourname on two datasets: IoT-Traffic and CIFAR-10, with varying values of $n$ of  10, 50, and 100.
For the \flame algorithm, we show the runtime of each component and overall system in Tab~\ref{tab:run_flame}. Note that the runtime of \ourname includes the time taken by the Kernel and data transfer between the host and the FPGA. For the IoT-Traffic dataset, we have achieved significant performance enhancement of 1 340.1, 382.1, and 288.9 times with $n$ being 10, 50, and 100, respectively. On the CIFAR-10 dataset, we obtained a speed-up of 513 and 506.7 for $n$ values of 10 and 50, respectively. 

Compared to the implementation without SMPC, FLAME requires approximately 2.62 seconds for 50 CIFAR10 models, whereas our approach only takes about 1.5 seconds, underscoring the superiority of using FPGAs for accelerating operations.

Our evaluation has demonstrated the impressive ability of our accelerators to effectively enhance performance across different datasets and varying values of $n$. 
The results mentioned above are obtained from a single FPGA. 
However, it is worth emphasizing that the computation of both Cosine Distance and Aggregation can be partitioned into multiple FPGAs and calculated simultaneously, thereby reducing latency.
These two components can account for up to $98\%$ of the total runtime. Therefore, running  $m$ FPGAs simultaneously can lead to a runtime reduction of almost $m$. This distributed FPGA computing approach offers a promising solution to further enhance the performance of our proposed framework.

\section{Related Work}
\label{sec:related}
\textbf{Privacy-Preserving Backdoor Defenses in FL.}
Baffle et al.~\cite{andreina2020baffle} requires clients to inspect wrong predictions of the aggregated model, making it compatible with SMPC. However, attackers can avoid detection by not changing not-triggered sample predictions. Trusted hardware approaches, such as poisoning defenses on TEEs, have been proposed but are impractical due to significant overhead~\cite{hashemi2021byzantine,mondal2021poster}. SMPC has been used for various approaches against poisoning attacks but has limited scalability due to expensive operations~\cite{nguyen22Flame,khazbak2020mlguard,dong2021flod}.\\
\textbf{FPGA Acceleration for FL.} Previous research studies have used FPGAs to speed up FL with HE for client privacy like the HE framework presented in~\cite{yang2020fpga} and the HW/SW co-design utilized in~\cite{wang2022pipefl}. However, neither of these works addresses backdoor attacks. Currently, no work has explored using FPGAs for SMPC acceleration.

\section{Conclusions}
\label{sec:conclude}

\noindent In this paper, we have presented \ourname, a framework that capitalizes on the benefits of FPGA-based computing to overcome performance bottlenecks inherent in software-only solutions. 
We demonstrate how FPGA-based TEEs can be leveraged to enable practical and privacy-preserving backdoor-aware FL aggregation on cloud FPGAs. \ourname offers more robust security guarantees than SMPC while minimizing the performance overhead. Its flexibility allows for the implementation of arbitrary aggregation schemes on secure FPGAs. Furthermore, our successful FPGA-accelerated implementation demonstrates the exceptional computational capabilities of FPGAs for accelerating FL algorithms.

\textbf{Acknowledgment.} This work is supported by the Deutsche Forschungsgemeinschaft (DFG, German Research Foundation) – SFB 1119 – 236615297, HMWK within the ATHENE project, the Hessian Ministery of Interior and Sport within the F-LION project, Intel as part of the Private AI Center, and Huawei as part of the OpenS3 Lab.

\bibliographystyle{IEEEtran}
\bibliography{mainbib}
\vfill

\end{document}